\documentclass{aastex}
\usepackage{emulateapj5}

\newcommand{\asca}{{\sl {ASCA\/}}}
\newcommand{\rxte}{{\sl {RXTE\/}}}
\newcommand{\euve}{{\sl {EUVE\/}}}
\newcommand{\cat}{{\sl {CAT\/}}}
\newcommand{\hegra}{{\sl {HEGRA\/}}}
\newcommand{\whipple}{{\it Whipple}}
\newcommand{\sax}{{\it BeppoSA$\!$X}}
\newcommand{\vfv}{$\nu F_{\nu}$}
\def\Ga{\Gamma}
\newcommand{\sy}{synchrotron}

\newcommand{\flux}{erg~cm$^{-2}$~s$^{-1}$}
\newcommand{\ep}{E_{\rm peak}}
\newcommand{\nh}{N_{\rm H}}
\newcommand{\nfnp}{\nu F_{\nu ,\rm peak}}
\newcommand{\nfn}{\nu F_{\nu}}

\slugcomment{to appear in ApJ}

\shorttitle{Tanihata et al.}
\shortauthors{Evolution of spectrum of Mrk~421}

\begin{document}


\title{Evolution of the Synchrotron Spectrum in Mrk~421
during the 1998 Campaign}


\author{Chiharu Tanihata\altaffilmark{1,2}, 
Jun Kataoka\altaffilmark{3}, 
Tadayuki Takahashi\altaffilmark{1,2}, and
Greg M. Madejski\altaffilmark{4}
}
\email{tanihata@astro.isas.ac.jp}


\altaffiltext{1}{Institute of Space and Astronautical Science,
        3-1-1 Yoshinodai, Sagamihara, 229-8510, Japan}
\altaffiltext{2}{Department of Physics, University of Tokyo,
        7-3-1 Hongo, Bunkyo-ku, Tokyo, 113-0033, Japan}
\altaffiltext{3}{Department of Physics, Tokyo Institute of Technology,
	Tokyo, 152-8551, Japan}
\altaffiltext{4}{Stanford Linear Accelerator Center, 
	Stanford, CA, 943099-4349, USA}


\begin{abstract}
The uninterrupted 7-day \asca\ observations of the TeV blazar Mrk~421
in 1998 have clearly revealed that X--ray flares occur repeatedly.  
In  this paper, we present the results of the time-resolved spectral 
analysis of the combined data taken by \asca, \rxte, \sax, and \euve.  
In this object -- and in many other TeV blazars -- the precise 
measurement of the shape of the X--ray spectrum, which reflects 
the high energy portion of the \sy\ component, is crucial 
in determining the high energy cutoff of the accelerated electrons 
in the jet.  
Thanks to the simultaneous broadband coverage,
we measured the 0.1--25 keV spectrum resolved 
on time scales as short as several hours,
providing a great opportunity to investigate the 
detailed spectral evolution at the flares.  
By analyzing the time subdivided observations,
we parameterize the evolution of the \sy\ peak, 
where the radiation power dominates, by fitting the 
combined spectra with a quadratic form
(where the $\nu F_\nu$ flux at the energy $E$ obeys 
$\log \nu F_\nu (E)=log(\nu F_{\nu,\rm peak}) - const \times 
(\log E - \log E_{\rm peak})^2 $).
In this case, we show that there is an overall trend 
that the peak energy $\ep$ and peak flux $\nfnp$ both increase or
decrease together. The relation of the two
parameters is best described as
$\ep\propto\nfnp^{0.7}$ for the 1998 campaign.
Similar results were derived for the 1997 observation,
while the relation gave a smaller index
when included both 1997 and 1998 data.
On the other hand, we show that this relation,
and also the detailed spectral variations,
differ from flare to flare within the 1998 campaign.
We suggest that the observed features are consistent
with the idea that flares are due to a appearance
of a new spectral component.
With the availability of the simultaneous TeV data, 
we also show that there exists a clear correlation between the
\sy\ peak flux and the TeV flux.
\end{abstract}




\section{INTRODUCTION}
The uninterrupted 7-day \asca\ observation of the TeV blazar Mrk~421
in 1998 has revealed that day-scale X--ray flares seen in previous 
observations were probably unresolved superpositions of many smaller 
flares \citep{tad00}.
The nearly--continuous observation allowed not only the
possibility to track the individual flares entirely from the rise 
to decay, but it also enabled quantitative statistical tests of 
the time series by employing the power spectrum or the structure 
function \citep{kataoka01,tanihata01}.  

The main characteristic of blazars is their high flux observed from 
radio to $\gamma$--rays coupled with strong variability and strong
polarization.  These properties are now successfully explained by the
scenario where blazars are active galactic nuclei (AGN), possessing 
jets aligned close to the line of sight, and accordingly the 
Doppler-boosted non-thermal emission from the jet dominates 
other emission components (e.g. Blandford \& K\"{o}nigl 1979; 
Urry \& Padovani 1995).  This is what makes blazars 
critical in understanding jets in AGN.  

The broadband spectra of blazars consist of two peaks, 
one in the radio to optical--UV range (and in some cases, reaching to
the X--ray band), and the other in the hard X--ray to $\gamma$--ray region.
The high polarization of the radio to optical emission suggests that 
the lower energy peak is produced via the synchrotron process by 
relativistic electrons in the jet.  The higher energy peak is 
believed to be due to Compton up-scattering of seed photons by 
the same population of relativistic electrons.  Several possibilities 
exist for the source of the seed photons;  these can be the
synchrotron photons internal to the jet \citep{jones74,ghisellini89}, 
but also external, such as from the broad emission line clouds
\citep{sikora94} or from the accretion disk \citep{dermer92,dermer93}.

The blazars with peak synchrotron output in the X--ray range 
also emit strongly in the $\gamma$--ray energies, and the 
brightest of those have been detected in the TeV range with 
ground-based Cherenkov arrays.  These are the so-called ``TeV blazars.''
In TeV blazars, variability of the synchrotron flux is measured to 
be the strongest and most rapid in the X--ray band, and thus it 
provides the best opportunity to study the electrons that are 
accelerated to the highest energies.  In particular, the \sy\ peak 
is a very important observable in two aspects: first because the 
flux at the peak represents the total emitted power from the blazar, 
and secondly because the peak frequency reflects the maximum energy  
of radiating particles gained in the acceleration process.

Mrk~421 is among the closest known blazars, at redshift of 0.031,
and was the first blazar (and also the first 
extragalactic source) discovered to be a TeV emitter.
It was first detected as a weak source by EGRET 
\citep{lin92}, and 9 months later, \whipple\
detected a clear signal from this object 
between 0.5 and 1.5 TeV \citep{punch92,petry96}.
Flux variability on various time scales has been observed,
including a very short flare with a duration 
of $\sim$1 hour \citep{gaidos96}.
Ever since, Mrk~421 has been repeatedly 
confirmed to be a TeV source by various ground-based telescopes.
It has also been one of the most studied blazars,
and have been a target of several multi-wavelength
campaigns \citep{macomb95,maraschi99,tad00}.

The multi-wavelength campaign of Mrk~421 in 1998 was one of the 
first opportunities to observe a blazar in the TeV range using several 
telescopes located in different locations in the world, 
so as to have as continuous coverage as possible \citep{felix99_421,piron01}.
Observations in other frequencies included 
X--ray observations by \asca, \rxte\ \citep{tad00}, 
and \sax\ (Maraschi et al.\ 1999; Fossati et al.\ 2000a,b),
EUV observations by \euve, optical observation with BVRI filters
organized by the WEBT collaboration
(http://www.to.astro.it/blazars/webt/),
and radio observations at the Mets\"{a}hovi Radio Observatory. 

In this paper, we present the results of the spectral analysis of 
X--ray and EUV data during the 1998 April campaign.
In particular, we estimate the location of the
\sy\ peak in the \vfv\ spectrum. 
So far, quantitative studies of the variation of 
the \sy\ peak have been conducted only for the two brightest 
blazars, Mrk~421 and Mrk~501.  
The largest variation of the \sy\ peak energy
was observed in Mrk~501 by \sax, where the peak energy 
shifted $\sim$2 orders of magnitude \citep{pian98}.
Collecting data from different epochs (separated by as long as years), 
\citet{fab01} have shown the relation of the form 
$\ep\propto F_{0.1-100 \rm keV}^n$ with $n\geq$2.
A similar analysis was done by \citet{fossati00b}
for the Mrk~421 data obtained by \sax\ in 1997 and 1998,
which showed a relation of 
$\ep\propto F_{0.1-10 \rm keV}^n$ with n=0.55.
In this paper we describe a continuous 7-day variation of the
\sy\ peak, which allows us to investigate 
the dynamical change of the \sy\ spectrum during the flares. 

We first describe the observations and data analysis in \S2.  
The results of spectral analysis are described in \S3;   
a summary of the results and a discussion of our findings 
with the emphasis on the temporal evolution of the spectrum 
are presented in \S4.

\section{OBSERVATION AND DATA ANALYSIS}
The week-long multi-frequency campaign of Mrk~421 was carried 
out in April 1998 \citep{tad00}.  The goal of this paper is to study
the evolution of the high energy \sy\ spectrum 
as a function of time during the campaign, 
and also the correlation with the TeV flux.  
This requires the best possible spectral coverage
in the X--ray range, for which we assembled all available data 
collected during the campaign, including: pointings by X--ray 
observatories \asca, \rxte, and \sax, as well as the EUV observations 
by \euve.

\subsection{\asca}
Mrk~421 was observed with \asca\ for 7 days continuously (except 
for the Earth occultation) 
during 1998 April 23.97 -- 30.8 UT (PI: T.~Takahashi).  
The \asca\ detectors -- 
two SISs (Solid-state Imaging Spectrometers; 
Burke et al.\ 1991; Yamashita et al.\ 1997)
and two GISs (Gas Imaging Spectrometers; Ohashi et al.\ 1996)
were in operation.  We refer to \citet{tanihata01} to the
details of the observation modes and data reduction.
The obtained spectra were rebinned such that all bins 
have the same statistics. 

Since late 1994, the efficiency of both SIS detectors below 
$\sim$1 keV has been decreasing over time.  The problem is believed 
to be caused by the increased residual dark current level,
and also by the decrease in the charge transfer efficiency,
although effort is still underway to model the effects.
While it has been reported that it may be possible to 
parameterize the degradation by an additional absorption column 
as a function of time \citep{yaqoob00}, since our observation 
concerns the precise shape of the continuum spectrum, we will use 
only the data obtained by the GIS for spectral analysis.
(Note that the X--ray spectra in Figure~2 of \citet{tad00}
show the data obtained by SIS. For this, there is a too strong 
decrease of inferred flux at the lower energy, which is due to 
instrumental effects described above.)

\subsection{\rxte}
The \rxte\ observations (PI: G.~Madejski) were mainly 
coordinated to coincide with
the \asca\ observation period, but were conducted mostly only during the 
low-background orbits.  The \rxte\ extends the \asca\ bandpass to a 
higher energy with the instruments PCA (Proportional Counter Array; 
2--60 keV) and HEXTE (High Energy X--ray Timing Experiment; 15--200 keV).
However, because both the flux from the source and the sensitivities 
of the detectors decrease towards higher energies, in the analysis
below we use only the data from the PCA, in the energy range of 3--25 keV.

All data reduction was performed using the HEASOFT software packages, 
based on the REX script, provided by the \rxte\ Guest Observer
Facility (GOF).  The screening criteria excluded the data when 
the elevation angle from the Earth's limb was lower than 10\degr, 
and the data acquired during the SAA passages.  
Since the PCUs are activated during the SAA passage, we used 
the data only after when the background level dropped to the 
quiescent level and became stable, which is typically 30 minutes 
after the peak of a SAA passage.

For the best signal-to-noise, we selected events only from the top 
xenon layer, and excluded events from other gas layers.  For the maximum 
consistency among the observations, we used only the three detectors 
which were turned on for all the observations (known as PCU0, PCU1, 
and PCU2).  Since the PCA is not an imaging detector, the blank-sky 
observations must be used to estimate the background spectrum.
For this, we used the background models provided by the \rxte\ GOF
to evaluate the background during each observation.
We note that the residual background contamination of the data 
within the energy bandpass selected by us was always significantly lower
than the source counts.  The mean net \rxte\ PCA count rate from Mrk~421 
in the 10 - 20 keV range for 3 PCUs as above was at least 
$\sim 3$ counts s$^{-1}$ while the total residuals 
from blank-sky in this energy range ${\sl plus}$ 
the fluctuations of the Cosmic X-ray Background were lower 
than 0.2 counts s$^{-1}$ (see HEASARC Web page regarding the \rxte\ PCA 
background), providing the confidence that 
an imprecise background subtraction did not affect our results.  

\subsection{\sax}
The \sax\ observations (PI: L.~Chiappetti) started 3 days before 
the \asca\ observations 
(Maraschi et al.\ 1999; Fossati et al.\ 2000a,b).  We use the data 
obtained by the LECS (Low Energy Concentrator Spectrometer; 0.1--10 keV), 
which extends the bandpass to lower energy, and also the
MECS (Medium Energy Concentrator Spectrometer; 1.3--10 keV). 
The data analysis was based on the linearized event files, together 
with the appropriate background event files and response matrices, 
provided by the \sax\ Science Data Center (SDC).  Data
reduction was performed using the standard HEASOFT software, XSELECT.

We extracted the source photons from a circular region centered at 
the source with a radius of 8 arcmin for both LECS and MECS.
For the background photons, we constructed the spectra by 
extracting the photons from the same detector regions 
using blank-sky observations.  
Each spectrum was rebinned using the grouping templates
provided by the \sax\ SDC, and we limited the energy
range to 0.12--4.0 keV for LECS, and 1.65--10.5 keV 
for MECS, as suggested in the \citet{saxcookbook}.

\subsection{\euve}
The \euve\ observations
covered the entire campaign of Mrk~421 from April 19.4 to May 1.2
(PI: T.~Takahashi).  The source photons were extracted in a 
12 arcmin aperture.  In order to estimate the absolute flux 
in the EUV range, we normalized it to be consistent with the 
simultaneous data obtained by the other observatories.  Given 
that the EUV spectrum is difficult to determine within the 
rather limited bandpass, we used only the observed integrated 
count rate (cf. Marshall, Fruscione \& Carone 1995).  For this, 
we used the period where the observations by all observatories 
overlapped (MJD 50927.03 - 50927.34; \S\ref{sec:combine}).  
During this period, the average \euve\ count rate was 
0.46$\pm$0.01 cts/s. 
At the same time, the 0.13--0.18 keV flux derived from 
the combined LECS, MECS, GIS, and PCA spectrum gives 
$\nu F_{\nu} = (1.78 \pm 0.04) \times 10^{-10}$ \flux, where the error 
represents the statistical error. 

An additional uncertainty arises from the fact that the
difference in the spectral shape is not taken in to account.
This can be estimated from the results in \citet{marshall95}.
Taking $\nh=1.5 \times 10^{20}$ cm$^{-2}$ for Mrk~421, 
assuming that
the spectrum index varies from 0.5 to 1.0, the flux density
would vary from 1363--1394 $\mu$Jy for 1 cts/s in the \euve\ range.
This gives an additional 2.5\% error as a systematic error.

In the following analysis, we use the above value for scaling 
the observed \euve\ count rate to flux.  Thus,
\begin{equation}
\nu F_{\nu} ({\rm EUVE}) = 1.78\times 10^{-10}
         \left( \frac{F_{\rm cts/s}}{0.46} \right)  
{\rm erg\,s}^{-1} {\rm cm}^{-2},
\end{equation}
where $F_{\rm cts/s}$ is the \euve\ count rate.  

\section{RESULTS}
\subsection{The Shape of the 0.1--25 keV Spectrum}
\label{sec:combine}
During April 24, observations by all 3 X--ray satellites, 
\asca, \sax, \rxte, and also \euve\ happened to exactly overlap 
continuously for 26.5 ksec (MJD 50927.03 - 50927.34).  This provides 
us with high quality data over a very broad energy coverage,
spanning more than 2 decades, from 0.12 to 25 keV.  The net exposure
for each instrument during the overlapping 26.5 ks period were
9.1, 13.3, 3.6, and 5.5 ksec for GIS, PCA, LECS, MECS, respectively.

We first fitted the spectrum for each instrument with 
a power-law model, with a low energy absorption
fixed to the Galactic value, $\nh = 1.5\times 10^{20}$ cm$^{-2}$
\citep{elvis89}. 
The results are shown in the top half of Table~\ref{tbl:fit},
suggesting that the fit is not adequate for 
any of the individual instruments.
This certainly suggests
that a more complex model, rather than just a simple power-law
function, is required to describe the spectrum.

The shape of the \sy\ spectrum is determined from the
energy distribution of the radiating electrons, which cuts off at 
the high energy representing the acceleration limit.
As described in the introduction, the high energy 
end of the \sy\ spectrum in TeV blazars is located in the 
X--ray range;  this is most likely the reason that the high-quality 
X--ray data cannot be described by a single power-law.

Given the curved spectrum for each instrument,
we fitted the combined X--ray spectrum with
various functions to model the shape of the observed curvature.
We note that our aim is not to develop the shape of the curvature from
first principles, but rather, to reproduce the observed data. 
For this reason, we considered the following models for comparison:  
(1) A broken power-law,
(2) a curved power-law,
(3) a power-law with an exponential cutoff, and 
(4) a quadratic function in $\log\nu$-$\log$\vfv\ plane.
The curved power-law that we define below is similar to the broken 
power-law model, but the change in the spectral index is not 
discrete.  The form attempted by us is 
$F(E)=K E^{-\Ga_1}(1+\frac{E}{E_{\rm br}})^{-\Ga_2}$.
The form for the exponentially cutoff power law is 
$F(E)=K E^{-\Ga_1} \exp(-E/E_{\rm br})$,
and the quadratic function is described as
$\log \nu F_\nu (E)=log(\nu F_{\nu ,\rm peak}) - 
a (\log E - \log E_{\rm peak})^2 $.  

Since there are slight differences in the cross-calibration 
between the instruments, we allow the normalization to vary 
for each instrument.  For all results below, the flux will 
be normalized to the GIS value, unless otherwise noted.  
The $N_{\rm H}$ is fixed to the Galactic value.  

By fitting the spectrum with the models above, we first found that 
the power-law with an exponential cutoff is not a good representation  
of the data:  the model cutoff is too sharp at high energies.
The broken power-law model is also not acceptable,
and the curved power-law is significantly better.  
We also find that the quadratic function fits the 
spectrum  surprisingly well.  
The fitted spectrum and the residuals to the quadratic model 
are shown in Figure~\ref{fig:spec}, where all LECS, MECS, GIS, and PCA
spectra are shown to be well described with this single function.
The results of the spectral fits by the 4 models 
are summarized in in the bottom half of Table~\ref{tbl:fit}.

\subsection{The Time Resolved Spectra}
The \euve\ observation covered the whole campaign (April 19.4 to May 1.2), 
and thus completely overlapped our 7-day \asca\ observation.
This adds a data point at 0.13-0.18 keV in addition to the 
\asca\ spectrum in the 0.7--7 keV range, providing us the best opportunity 
to study the continuous spectral variability.  In fact this is the longest 
continuous monitoring of this object so far spanning over the 
0.13--7 keV range.

We divided the whole observation into shorter segments with each 
having a duration of 10 ksec.  
In order to combine the GIS and \euve\ data,
the GIS count rate spectra must first be 
converted to the source flux spectra
as the \euve\ count rate data converts directly into flux (\S2.4).
For this, we first fit the GIS spectrum with 
a curved power-law model with Galactic absorption. This gives 
the source spectrum in the GIS energy range.  Note that this 
procedure depends on the selected model, but 
long as it describes the spectrum well, this should introduce 
the smallest bias.  We then combine this with the \euve\ 
data, which gives the combined source spectrum.

In fitting the combined \euve/GIS spectra, 
we first started with a single power-law model.  
In similarity to the results in \S 3.1, none of the 56 spectra
could be fit with a single power-law function, with their
average $\chi^2_\nu$ of 7.7, implying the curvature of the
spectra for all data sets.  Given that a quadratic model fits 
the spectrum equally well as the curved power-law model, 
but with fewer parameters (see \S3.1), 
we will assume the quadratic model for all data sets hereafter.  
In fact, because of the energy gap between the \euve\ and GIS, 
the fit with a curved power-law model resulted in having 
multiple minima in the $\chi^2$ plane (since the parameters are 
correlated), and the exact value of the best-fit 
parameters depended highly on the initial value. 

We checked whether the fit was adequate for all data segments.
The $\chi^2_\nu$ for each fit (shown in the bottom 
panel in Figure~\ref{fig:lcpeak}) ranged from 0.71 to 1.45. 
The number of degrees of freedom ranges from 60 to 187, 
which gave $P(\chi^2)>$5\% for 53 out of the total of 56 data points. 
For the other 3 data points (corresponding to times 230--240,
270--280, and 450--460 ksec), we found several line or edge-like 
structures in the spectra, which resulted in the increase of 
$\chi^2_\nu$. Although it is not clear whether this is due to some 
systematic effect or if it is a true property of the source, 
since our aim is to model the continuum spectrum correctly, we
excluded the energy ranges where those features are present. 
This improved the fit significantly, which is shown as the crosses.
With this, all the 56 spectra were well described with
the quadratic function.

The time evolution of the derived parameters in the fit
is shown in the top 3 panels in Figure~\ref{fig:lcpeak};
the peak flux, peak energy, and the curvature parameter $a$
(see definition in Table~\ref{tbl:fit}).  The peak energy is 
observed to shift between 0.5 keV and 1.2 keV, in which a general 
trend is found such that the peak energy is higher when the peak flux 
is higher. 
This is demonstrated as the filled circles in Figure~\ref{fig:fpep-all}.
The fit results are listed in Table~\ref{tbl:lcpeak}.

We do note that the actual validity of using a 
quadratic function in the entire 
0.1--25 keV range was checked in detail (including an 
examination of trends in residuals) only for the epoch when the 
overlapping \sax, \asca, and \rxte\ observations were available.  
Since we measured only the \euve\ flux below 0.7 keV in the
combined \euve/GIS fit,
the derived peak location can depend on the model function we assume. 
For instance if $any$ function was allowed, the peak can be
located at any energy within the gap.
For this, we emphasize that the derived peak location is the
estimated peak, assuming that the spectrum gap can be
extrapolated with the same quadratic function, and also that the
error bar describes only the statistical error.

We also analyzed the \sax\ data during the campaign.  Furthermore, in 
order to make a comparison with data obtained from another campaign, 
we also analyzed 
the \sax\ data obtained in 1997 (PI: G.~Vacanti, L.~Chiappetti).
We divided the observation into 
segments of the same 10 ks duration, and fit the 
combined LECS/MECS spectra with a quadratic function.  For the 
\sax\ data, we also fit the spectra with a curved power-law model 
for comparison.  
The derived peak flux and peak energy from 
both models are listed in Table~\ref{tbl:lcpeak_sax}.
In fitting the combined LECS/MECS spectra, 
the normalizations are allowed to vary independently,
and the values in Table~\ref{tbl:lcpeak_sax} are 
normalized to the MECS2 value.
The fitting results show that 
although the spectral fit parameter
is one less in the quadratic function,
both models equally represent the observed spectra.
In contrast to the combined \euve/GIS data,
there is no energy gap in the \sax\ data, which provides
support that our assumed quadratic form 
is appropriate in estimating 
the peak location in the \euve/GIS case.
The derived peak flux and peak energy using
both functions are similar, 
but we also note that the quadratic function 
gives a slightly lower peak energy than the
curved power-law function.

The derived peak flux and peak energy assuming the
quadratic model for the \sax\ observations are shown 
in Figure~\ref{fig:fpep-all} as the open circles (for 1998) 
and squares (for 1997).
Note that the \sax\ data points in Figure~\ref{fig:fpep-all}
are multiplied by a normalization factor of 1.07, 
which we adopted from the combined fit in \S3.1
It can be seen that the results during the 1998 campaign
lie on a similar slope as the results derived from the 
\euve/GIS data during the same campaign.

The relation of peak flux $\nfnp$ and peak energy $\ep$ for the 
\euve/GIS data set can be best described with a power-law function 
of $\ep \propto \nfnp^\alpha$, where 
$\alpha$=0.72$\pm$0.02 (or $\epsilon$=0.76$\pm$0.02 
in form $\ep \propto F_{\rm 0.1-10 keV}^\epsilon$; 
see description below).
By including also the \sax\ data points during 1998, the index becomes
$\alpha$=0.77$\pm$0.02 ($\epsilon$=0.72$\pm$0.02).  
It can be seen that the data from 1997 lie on a somewhat different slope.
The relation derived from the 1997 data {\sl alone} gives
$\alpha$=0.96$\pm$0.09 ($\epsilon$=0.79$\pm$0.06), 
slightly steeper than the index derived from the 1998 campaign.  
Fitting all data together we obtain 
$\alpha$=0.59$\pm$0.01 ($\epsilon$=0.50$\pm$0.01).
All errors correspond to a 1$\sigma$ error.

A similar analysis was performed by \citet{fossati00b} for the 1997 
and 1998 \sax\ data.  \citet{fossati00b} note a tight 
relationship between the measured integrated
0.1-10 keV flux $F_{\rm 0.1-10 keV}$ and 
the peak energy $\ep$ such that 
$\ep \propto F_{\rm 0.1-10 keV}^\epsilon$ with $\epsilon$=0.55$\pm$0.05,
similar to the index derived from our results including all data sets.
For a direct comparison, we fit simultaneously both 1997 and 1998 
for \sax\ data points only.  This gave
$\alpha$=0.52$\pm$0.02 ($\epsilon$=0.45$\pm$0.01). 
The estimated $\epsilon$ is slightly smaller than 
reported in \citet{fossati00b}, which might be affected by 
different subdividing of the observations.  
Nonetheless, it is worth noting that 
the slope derived from the combined data appears be 
softer (smaller $\alpha$ or $\epsilon$) 
than each epoch considered alone.

Furthermore, the relation appeared to also vary between individual 
flares during the 1998 campaign.  In some flares the peak energy 
clearly increases together with flux, while the correlation 
is weaker in others.  This can be seen clearly in 
Figure~\ref{fig:fpep-each}, which shows the evolution of the 
synchrotron peak in each flare.  The peak energy has a trend 
to shift to the higher energy and then return to the initial 
value together with the flux, but there is no common slope.  

\subsection{Spectral Variations at Flares}
In the previous section, we described the spectral variations
as the evolution of the location of the peak in the \vfv\ spectrum.
Here, we take advantage of the wide energy range
coverage afforded by the use of by \euve, \asca, and \rxte\ data
to investigate spectral variations in more detail.  
Given that the \rxte\ observations happen mostly during the
flare minima rather than at their peaks (see Figure~1 in 
Takahashi et al.~2000), here we focus on the beginning (rising portion) 
of the flares.  

We divided the observations into 10 ksec segments, similar
to the previous section.  For each segment, the GIS and PCA spectra 
are fitted simultaneously with a curved power-law function 
with an additional constant factor for the normalization. 
The derived source spectra in the GIS and PCA range 
are combined with the \euve\ data point, which gives the total 
source spectrum in the energy range of $\sim$0.1--25 keV.  
In the top panels of 
Figures~\ref{fig:vfvratio8}--\ref{fig:vfvratio34}, 
we show the evolution of the \vfv\ spectrum derived in this manner 
for the 2nd, 4th, and 5th flare (starting from 80 ksec, 260 ksec, 
and 340 ksec, respectively, as defined in Figure~\ref{fig:lcpeak}).
Note that the spectra are further rebinned for this Figure.  
For each flare, we also plot a subsequent spectrum, but those 
only include the \euve\ and GIS data. 

To see the evolution more clearly, we calculated the
ratio of the rebinned spectra by dividing these ``subsequent'' 
spectra by the spectrum measured at the onset of a given flare 
(integrated over 80--90, 260--270, 340--350 ksec, respectively). 
This is shown in the bottom panels of 
Figures~\ref{fig:vfvratio8}--\ref{fig:vfvratio34}.
Interestingly, these three flares all showed different features. 
It can be seen that Flare 2 starts with a hardening 
at the very highest energy range with a concave shape in the ratio.
As time progresses, the whole energy range becomes dominated by a new 
spectral slope.  In contrast to this, in Flare 4 the lower energy 
photons appear to precede the total flux rise.  Flare 5 was 
similar to Flare 2, but all energy bands increase together, 
with a larger amplitude at higher energies.  We also studied 
the spectral variation for other flares, but a detailed 
investigation was rather difficult without a denser temporal 
coverage by \rxte.  

\subsection{The X--ray and TeV Correlation}
Since X--ray and TeV energies both reflect the electrons accelerated 
to the highest energies in TeV blazars, simultaneous X--ray and TeV 
observations provide an important tool to probe the emission region 
of the jet (e.g., Krawczynski et al.\ 2002, and references therein). 
On the other hand, since blazars are highly variable on short time scales, 
exactly simultaneous observations are essential.  Thanks to the 
relatively high brightness of the source during the 1998 campaign, it 
was possible to generate a significant measurement for the separate 
TeV pointings every night (Aharonian et al.\ 1999; Takahashi
et al.\ 2000).  Here we provide the X--ray spectrum from the exactly 
simultaneous time regions with the TeV observations.

For each TeV observation by \hegra\ \citep{felix99_421} and \whipple\ 
\citep{tad00}, we collected all existing X--ray data  during the 
TeV exposure, from the \asca, \rxte, and \euve\ observations
(the \cat\ observations mostly overlap with the \hegra\ observations).
In particular, all \whipple\ pointings during the \asca\ observations 
happened to coincide with the \rxte\ observations, providing an X--ray
spectrum ranging from 0.1--25 keV, simultaneous with the TeV observation.

We generated the combined X--ray spectrum using the same procedures as 
described in the previous sections.  We then fitted the X--ray spectrum 
for each period assuming a quadratic function.  The X--ray spectra during 
the \hegra\ observations (consisting of the \euve\ and \asca\ data) 
are shown in Figure~\ref{fig:vfv_x}(a).  The spectrum for each 
observation is shown as different symbols, and the dashed line shows 
the modeled quadratic function derived via the fit.
The X--ray spectra during the \whipple\ observations,
which consist of the \euve, \asca\ and \rxte\ data
are shown in Figure~\ref{fig:vfv_x}(b).  Note that the spectra are 
further rebinned in the figures.  While each integration time is as 
short as several hours, it can be seen that the curved shape of the 
high energy peak of the \sy\ spectrum is clearly resolved, and it is 
evident that the peak is located in the X--ray range.  

The results of fits to each X--ray spectrum concurrent with the TeV
pointings are summarized in Table~\ref{tbl:xtev}.  
We also list the simultaneous TeV fluxes, in which
the \hegra\ data were taken by \citet{felix99_421}.
In Figure~\ref{fig:xtev}, we show the correlation 
of the derived location of the \sy\ 
peak flux with the TeV flux.  Since the sensitive energy ranges 
of the two TeV telescopes are slightly different, we used the 
integrated flux in Crab units. A good correlation can be seen, showing 
that the TeV flux is higher when the \sy\ peak flux is higher.
A best fit to the relation with a power-law function gives 
$F_{\rm TeV} \propto F_{\rm sy,peak}^{\alpha}$ with
$\alpha$=1.7$\pm$0.3.  

\section{SUMMARY AND DISCUSSION}
We have performed detailed analysis of the combined \euve, \asca, \sax, 
and \rxte\ data collected during the long look campaign of Mrk~421 
in 1998, resulting in the measurement of the time-resolved spectrum
in a broad energy range of 0.1--25 keV in segments as short as several 
ksec.  These are among the highest quality spectra with regard to both
photon statistics and energy coverage so far for any blazar, providing 
the precise spectral shape at the highest energy end of the \sy\ 
spectrum.  

We have shown that both curved power-law function
and quadratic function in 
$\log\nu$--$\log$\vfv\ space reproduce the combined
\euve, \sax, \asca, and \rxte\ spectrum.
For the epochs  where \sax\ data do not exist,
we assumed that the spectral gap between 
$\euve$ and $\asca$ can be extrapolated with a
quadratic function, and we showed that the energy and the flux  
at the \sy\ peak in the \vfv\ spectrum are correlated, showing an 
overall trend of a higher \sy\ peak energy for higher peak flux, 
but the details of this relation differed from flare to flare. 
In particular, within the three flare beginnings, one flare started 
from a rise in the higher energy while another started from a 
rise in the lower energy.  The relative amplitude of the rise 
in different energies also differed from flare to flare.
An interesting result was shown from the 2nd flare,
where it was indicated that this flare started with a hardening 
at the higher energy. This feature is difficult to describe 
by a simple cooling or acceleration of a single electron distribution
(see e.g. Kataoka 2000), and thus this concave curvature indicates that
two different spectral components were likely to co-exist at the 
beginning of the flare.  We remark that such a behavior is observed 
in only one flare, but the importance is that we observed a flare 
that could not be explained by a single electron distribution.  
This suggests an appearance of a new component, which generated the 
observed flare.

Recently, many theoretical studies has been performed to model the 
energy dependence, or time lags observed in the day-scale flares in 
blazars.  One approach considers models with a homogeneous emission 
region \citep{masti97, chiaberge99, Li00}.  For instance, \citet{masti97} 
have suggested that a sudden increase in the maximum energy of the 
accelerated electrons could result in the X--ray and TeV flare 
observed in Mrk~421.  Similar results were shown by \citet{kataoka00},
in reproducing the soft-lag observed in PKS~2155--304.  
However, most of these models assume a change in some parameter 
in the emission region (such as the magnetic field, maximum energy of 
electrons, number density of electrons, etc.).  The observational results 
for Mrk~421 described above lead us to suggest a different scenario.

An alternative scenario has each flare forming as a result of 
a separate electron distribution.  This was suggested in modeling 
the high state SED of Mrk~501 \citep{kataoka99}.  They applied the 
one-zone homogeneous synchrotron self-Compton model to the SED 
from April 1997, and concluded that a single electron distribution 
is insufficient to reproduce the observed synchrotron spectrum.  
This scenario could work for instance when the jet is emitted
intermittently from the central engine, and several
separate regions generate different emission components. 
Another viable scenario can be provided by an internal shock model, 
where the light curve results from a superposition of
many flares due to the collisions of shells,
which may occur when a faster shell catches up to a slower shell
(Ghisellini 2001, Spada et al.\ 2001; Sikora et al.\ 2001)

In fact, \citet{tanihata02} has shown from simulations that 
the internal shock model can naturally explain various variability 
properties observed in TeV blazars.  One observational fact is that 
the X--ray flares always appear to lie on top of an underlying offset-like 
component.  In the internal shock model, flares can be considered 
as arising from collisions of shells that had the largest difference 
in the initial velocity.  All other collisions generate smaller 
amplitude, longer flares, which pile up to generate the offset component.  
In this case, it was shown that the flares will tend to have a higher 
\sy\ peak energy as compared with that of the offset component. 

Assuming that flares are due to an emergence of a new component
with a higher \sy\ peak energy than the pre-existing component, the 
\sy\ peak energy in the observed spectrum would appear as shifting 
to a higher energy.   It is interesting to note that our
observations are consistent with this, with the overall trend of 
the \sy\ peak energy being higher during flares.
Furthermore in this case, the spectral evolution would 
depend on the relation of the new and pre-existing spectra.
This would naturally explain the different spectral evolution
among different flares.

Finally, we remark again that due to the data gap 
between the \euve\ and \asca\ data in our observation,
the results concerning the spectrum peak
are somewhat tentative. Because of the rapid 
variability of the source in short time scales,
the effective area of the instrument is most important.
Long-look observations by the new observatories such as 
$XMM$-$Newton$ should provide data allowing a more detailed 
analysis of the flaring mechanism, leading to the dynamics of the
accelerated electrons in blazar jets.

\acknowledgements
We thank the anonymous referee for numbers of 
constructive comments to improve the paper.
Support for this work was provided by the Fellowship of
Japan Society for Promotion of Science for Young Scientists, 
and by the Department of Energy contract to SLAC no. DE-AC3-76SF00515.    
We also acknowledge the NASA Chandra grants via SAO grant no. GO1-2113X.





\figcaption[]{
The joint fit with a quadratic function to the LECS, MECS, GIS, and PCA
spectra during the period
when the \asca, \sax, and \rxte\ observations overlapped
(MJD 50927.03 -- 50927.34).
The top panel shows the photon spectrum for each 
instrument. The solid line shows the best-fit model.
The bottom four panels shows the residuals of 
the observed data to the model.  See Table~\ref{tbl:fit} for 
the values of the fit parameters.  
\label{fig:spec}
}

\figcaption[]{The time evolution of the 
spectrum fit parameters to the 
combined \euve\ and \asca\ spectrum,
assuming a quadratic function (see text).
The top three panels 
show the peak flux (top), peak energy (second), 
and the curvature parameter (third; see definition in Table~\ref{tbl:fit}).
The flux is in units of 10$^{-12}$ \flux, and peak energy is in keV.
The bottom panel shows the reduced-$\chi^2$ for each of the fits.
Three out of the 56 spectra showed a poor fit ($P(\chi^2)<$5\%)
due to several line or edge-like structures, where the fit 
is significantly improved after excluding the energy ranges
that included these structures, shown as the crosses.
\label{fig:lcpeak}
}

\figcaption[]{The correlation of the 
\sy\ peak flux and peak energy
for Mrk~421, assuming a quadratic function.
Results derived from the combined \euve\ and \asca\ spectra
during the 1998 campaign are shown as the filled circles.
Results from the \sax\ spectra from the 1998 campaign
and 1997 observation are shown as the open circles and
squares, respectively.
The fluxes for the \sax\ data are multiplied by a factor of 
1.07 to normalize them to the \asca\ fluxes.
Each data point corresponds to the value derived from a spectrum 
integrated over 10 ksec.
\label{fig:fpep-all}
}
\figcaption[]{The time evolution of the correlation of the 
\sy\ peak flux $\nfnp$ and peak energy $\ep$
for Mrk~421 during the 1998 campaign,
plotted separately for each of the flares defined 
in Figure~\ref{fig:lcpeak}.
The initial values in each period are shown as a star,
and the connecting line shows the time order.
This Figure illustrates that the relation of $\nfnp$ and $\ep$ 
differ from flare to flare.
\label{fig:fpep-each}
}

\figcaption[]{
The spectral evolution 
during the rise of the second flare in the Mrk~421 
observation during the 1998 campaign.
The ratio of the spectrum to that of the onset 
of the rise is shown in the lower panel. 
It can be seen that the flare starts with
a hardening at the highest energy range
with a concave shape in the ratio,
which with time, propagates down to the lower energies.
In the last spectrum, which is 40 ks after the initial one, 
the whole spectrum is described with the new slope.
\label{fig:vfvratio8}
}

\figcaption[]{
The spectral evolution 
during the rise of the fourth flare in the Mrk~421 
observation during the 1998 campaign.
The ratio of the spectrum to that of the onset 
of the rise is shown in the lower panel.
\label{fig:vfvratio26}
}
\figcaption[]{
The spectral evolution 
during the rise of the fifth flare in the Mrk~421 
observation during the 1998 campaign.
The ratio of the spectrum to that of the onset 
of the rise is shown in the lower panel. 
\label{fig:vfvratio34}
}

\figcaption[]{ 
The X--ray spectra of Mrk~421 obtained simultaneously with the
(a) \hegra\ and (b) \whipple\ observations during the 1998 campaign. 
The X--ray spectra with different symbols
correspond to each TeV observation shown in the label.
The X--ray data include those collected by the \euve\ and \asca\ satellites
for the \hegra\ pointing, and additionally, by the 
\rxte\ satellite for the \whipple\ observations, and clearly show 
the curvature of the synchrotron spectrum.  
\label{fig:vfv_x}
}

\figcaption[]{
The correlation of the synchrotron peak flux  and the TeV flux (in Crab units)
for the simultaneous observations of Mrk~421 during the 
campaign in 1998.  The \sy\ peak flux is derived from observations by 
\euve\ and \asca\ (and \rxte\ for the \whipple\ pointings)
during the exactly simultaneous epochs where the TeV data are available.
The letters H and W in the legend denote \hegra\ and \whipple\ pointings.
\label{fig:xtev}
}

%
\begin{deluxetable}{ll ccc ccc c}
\tablecaption{Results of Spectral Fits for Mrk~421
during MJD 50927.03 -- 50927.34
\label{tbl:fit}}
\tablewidth{0pt}
\tablecolumns{9} 
\tabletypesize{\scriptsize}
\tablehead{
Instrument\tablenotemark{a}  & Model  & $\Ga_1$\tablenotemark{b} &
$\Ga_2$\tablenotemark{b} &$E_{\rm br}$\tablenotemark{c} &  $E_{\rm peak}$\tablenotemark{c} & $\nfnp$\tablenotemark{d} & $a$\tablenotemark{g} & $\chi^2_{\nu}$ (dof) 
}
\startdata
LECS (0.12-4) &   PL & 2.20$\pm$0.01 &-&-&-&-&-& 2.85 (83)  \\
MECS (1.65-10.5)& PL & 2.76$\pm$0.02 &-&-&-&-&-& 1.28 (193)  \\
GIS (0.7-7) &     PL & 2.52$\pm$0.01 &-&-&-&-&-& 1.29 (124)  \\
PCA (3-25) &      PL & 2.91$\pm$0.01 &-&-&-&-&-& 2.63 (50)  \\
\hline
LECS,MECS,GIS,PCA & broken PL & 2.21$\pm$0.01 &2.90$\pm$0.01 &2.24$\pm$0.04 &-&-&-& 1.86 (451) \\
LECS,MECS,GIS,PCA & curved PL\tablenotemark{e} & 1.77$\pm$0.04 & 1.66$\pm$0.03 & 2.5$\pm$0.3 &-&-&-& 1.08 (451) \\
LECS,MECS,GIS,PCA & cutoff PL\tablenotemark{f} & 2.17$\pm$0.01 &-& 7.80$\pm$0.14 &-&-&-& 1.74 (452) \\
LECS,MECS,GIS,PCA & quadratic\tablenotemark{g} &-&-&-& 0.47$\pm$0.02 & 217$\pm$3 & 0.42$\pm$0.01 &1.07 (452) \\
\enddata
\tablenotetext{a}{The numbers in parentheses show the energy range used 
	in the spectral fitting}
\tablenotetext{b}{Photon index}
\tablenotetext{c}{in keV}
\tablenotetext{d}{in $10^{-12}$\flux}
\tablenotetext{e}{$F(E)=K E^{-\Ga_1}(1+\frac{E}{E_{\rm br}})^{-\Ga_2}$}
\tablenotetext{f}{$F(E)=K E^{-\Ga_1} \exp(-E/E_{\rm br})$}
\tablenotetext{g}{$\log \nfn (E)=\log (\nfnp) - a (\log E - \log \ep)^2$}
\tablecomments{All errors represent the 1$\sigma$ error.}
\end{deluxetable}

\begin{deluxetable}{c| cccl }
\tablecaption{Spectral Fit Parameters
for the combined \euve\ /\asca\ spectra.
\label{tbl:lcpeak}}
\tablewidth{0pt}
\tablecolumns{9} 
\tabletypesize{\scriptsize}
\tablehead{
Observation\tablenotemark{a} 
 & $\ep$\tablenotemark{b} 
 & $\nfnp$\tablenotemark{c} 
 & $a$\tablenotemark{d} 
 & $\chi^2_{\nu}$ (dof) 
}
\startdata
 1 & 0.53$\pm$0.02 & 230$\pm$2 & 0.39$\pm$0.02 & 1.06 (66)\\
 2 & 0.51$\pm$0.02 & 244$\pm$3 & 0.39$\pm$0.02 & 0.95 (85)\\
 3 & 0.57$\pm$0.02 & 266$\pm$2 & 0.39$\pm$0.01 & 0.95 (122)\\
 4 & 0.69$\pm$0.02 & 280$\pm$2 & 0.48$\pm$0.01 & 1.22 (106)\\
 5 & 0.62$\pm$0.01 & 331$\pm$2 & 0.42$\pm$0.01 & 1.08 (116)\\
 6 & 0.72$\pm$0.02 & 372$\pm$2 & 0.40$\pm$0.01 & 0.97 (137)\\
 7 & 0.63$\pm$0.04 & 377$\pm$5 & 0.41$\pm$0.02 & 0.89 (145)\\
 8 & 0.60$\pm$0.03 & 323$\pm$4 & 0.46$\pm$0.02 & 1.20 (87)\\
 9 & 0.49$\pm$0.02 & 277$\pm$3 & 0.38$\pm$0.02 & 0.80 (82)\\
10 & 0.55$\pm$0.02 & 270$\pm$2 & 0.36$\pm$0.01 & 0.92 (79)\\
11 & 0.58$\pm$0.02 & 278$\pm$2 & 0.36$\pm$0.01 & 1.10 (137)\\
12 & 0.59$\pm$0.02 & 307$\pm$2 & 0.36$\pm$0.01 & 1.04 (126)\\
13 & 0.59$\pm$0.02 & 337$\pm$2 & 0.38$\pm$0.01 & 0.97 (125)\\
14 & 0.66$\pm$0.02 & 358$\pm$2 & 0.39$\pm$0.01 & 1.01 (154)\\
15 & 0.69$\pm$0.02 & 324$\pm$3 & 0.43$\pm$0.02 & 0.73 (97)\\
16 & 0.65$\pm$0.02 & 303$\pm$2 & 0.37$\pm$0.01 & 0.89 (87)\\
18 & 0.71$\pm$0.02 & 333$\pm$2 & 0.35$\pm$0.01 & 0.93 (98)\\
19 & 0.76$\pm$0.02 & 364$\pm$2 & 0.37$\pm$0.01 & 0.92 (173)\\
20 & 0.68$\pm$0.02 & 346$\pm$2 & 0.38$\pm$0.01 & 1.16 (144)\\
21 & 0.77$\pm$0.02 & 337$\pm$2 & 0.42$\pm$0.01 & 1.04 (132)\\
22 & 0.86$\pm$0.02 & 391$\pm$2 & 0.40$\pm$0.01 & 1.14 (169)\\
23 & 0.89$\pm$0.02 & 458$\pm$2 & 0.43$\pm$0.01 & 0.97 (184)\\
24 & 0.73$\pm$0.03 & 423$\pm$3 & 0.34$\pm$0.02 & 1.21 (119)\\
25 & 0.87$\pm$0.03 & 379$\pm$3 & 0.41$\pm$0.02 & 0.85 (113)\\
26 & 0.85$\pm$0.02 & 366$\pm$2 & 0.36$\pm$0.01 & 0.88 (95)\\
27 & 0.71$\pm$0.02 & 336$\pm$2 & 0.36$\pm$0.01 & 1.14 (153)\\
28 & 0.68$\pm$0.02 & 397$\pm$3 & 0.38$\pm$0.01 & 1.09 (146)\\
29 & 0.74$\pm$0.02 & 404$\pm$3 & 0.39$\pm$0.01 & 0.85 (125)\\
30 & 0.70$\pm$0.01 & 423$\pm$2 & 0.41$\pm$0.01 & 1.08 (176)\\
31 & 0.69$\pm$0.01 & 442$\pm$2 & 0.40$\pm$0.01 & 0.88 (181)\\
32 & 0.74$\pm$0.02 & 439$\pm$3 & 0.41$\pm$0.01 & 0.81 (124)\\
33 & 0.66$\pm$0.02 & 459$\pm$3 & 0.38$\pm$0.01 & 1.07 (136)\\
34 & 0.83$\pm$0.02 & 390$\pm$3 & 0.43$\pm$0.01 & 0.97 (93)\\
35 & 0.72$\pm$0.02 & 350$\pm$2 & 0.34$\pm$0.01 & 1.02 (135)\\
36 & 0.73$\pm$0.03 & 365$\pm$2 & 0.30$\pm$0.01 & 0.78 (136)\\
37 & 0.85$\pm$0.02 & 388$\pm$2 & 0.34$\pm$0.01 & 0.96 (165)\\
38 & 0.95$\pm$0.02 & 413$\pm$2 & 0.33$\pm$0.01 & 0.97 (187)\\
39 & 1.11$\pm$0.02 & 515$\pm$2 & 0.33$\pm$0.01 & 1.07 (182)\\
40 & 1.09$\pm$0.03 & 469$\pm$2 & 0.36$\pm$0.01 & 0.96 (146)\\
41 & 1.02$\pm$0.04 & 423$\pm$2 & 0.32$\pm$0.02 & 1.11 (153)\\
42 & 0.88$\pm$0.02 & 348$\pm$2 & 0.37$\pm$0.01 & 0.70 (92)\\
43 & 0.76$\pm$0.03 & 352$\pm$2 & 0.32$\pm$0.01 & 0.98 (131)\\
44 & 0.76$\pm$0.03 & 369$\pm$3 & 0.30$\pm$0.02 & 1.23 (84)\\
45 & 0.94$\pm$0.03 & 343$\pm$2 & 0.27$\pm$0.01 & 0.98 (97)\\
46 & 0.80$\pm$0.02 & 293$\pm$1 & 0.29$\pm$0.01 & 0.96 (126)\\
47 & 0.92$\pm$0.02 & 308$\pm$2 & 0.33$\pm$0.01 & 1.09 (138)\\
48 & 0.88$\pm$0.03 & 311$\pm$2 & 0.37$\pm$0.01 & 1.04 (130)\\
49 & 0.85$\pm$0.03 & 378$\pm$3 & 0.30$\pm$0.02 & 1.13 (80)\\
50 & 1.13$\pm$0.03 & 476$\pm$2 & 0.32$\pm$0.01 & 1.04 (132)\\
51 & 1.12$\pm$0.02 & 443$\pm$2 & 0.35$\pm$0.01 & 0.79 (144)\\
52 & 0.95$\pm$0.03 & 416$\pm$3 & 0.36$\pm$0.02 & 0.86 (112)\\
53 & 0.95$\pm$0.02 & 429$\pm$2 & 0.36$\pm$0.01 & 1.04 (171)\\
54 & 0.85$\pm$0.02 & 365$\pm$2 & 0.38$\pm$0.01 & 1.13 (180)\\
55 & 0.82$\pm$0.02 & 341$\pm$2 & 0.34$\pm$0.01 & 1.02 (152)\\
56 & 0.98$\pm$0.02 & 372$\pm$2 & 0.35$\pm$0.01 & 1.10 (157)\\
57 & 0.99$\pm$0.03 & 396$\pm$2 & 0.32$\pm$0.01 & 1.00 (166)\\
\enddata
\tablenotetext{a}{Each number represents the
observation divided into 10 ksec segments}
\tablenotetext{b}{in keV}
\tablenotetext{c}{in $10^{-12}$\flux}
\tablenotetext{d}{Definition is given in Table 1}
\tablecomments{All errors represent the 1$\sigma$ error.}
\end{deluxetable}

\begin{deluxetable}{ll |cccc | ccc}
\tablecaption{Spectral Fit Parameters for the 
\sax\ spectra
\label{tbl:lcpeak_sax}}
\tablewidth{0pt}
\tablecolumns{9} 
\tabletypesize{\scriptsize}
\tablehead{
Observation\tablenotemark{a} &  
 & \multicolumn{4}{c}{Quadratic}  
 & \multicolumn{3}{c}{Curved Power-law} \\
 & & $E_{\rm peak}$\tablenotemark{b} 
 & $\nu F_{\nu ,\rm peak}$\tablenotemark{c} 
 & $a$\tablenotemark{d} 
 & $\chi^2_{\nu}$ (dof) 
 & $E_{\rm peak}$\tablenotemark{b} 
 & $\nu F_{\nu ,\rm peak}$\tablenotemark{c} 
 & $\chi^2_{\nu}$ (dof) 
}
\startdata
1997.4.29 &  1 & 0.54$\pm$0.03 & 137$\pm$3 & 0.48$\pm$0.02 & 0.99 (126)	& 0.56$\pm$0.02 & 146$\pm$6 & 0.96 (125)\\
  &  2 & 0.46$\pm$0.02 & 124$\pm$2 & 0.48$\pm$0.02 & 1.21 (126)	& 0.49$\pm$0.02 & 125$\pm$4 & 1.23 (125)\\
&  3 & 0.50$\pm$0.01 & 131$\pm$2 & 0.46$\pm$0.01 & 1.16 (126)	& 0.53$\pm$0.01 & 134$\pm$3 & 1.15 (125)\\
\hline
1997.4.30 &  1 & 0.53$\pm$0.03 & 132$\pm$4 & 0.48$\pm$0.02 & 1.24 (126)	& 0.55$\pm$0.04 & 127$\pm$5 & 1.25 (125)\\
&  2 & 0.47$\pm$0.02 & 127$\pm$3 & 0.47$\pm$0.02 & 0.88 (126)	& 0.50$\pm$0.02 & 132$\pm$5 & 0.85 (125)\\
&  3 & 0.55$\pm$0.03 & 124$\pm$3 & 0.51$\pm$0.02 & 0.85 (126)	& 0.58$\pm$0.03 & 123$\pm$4 & 0.84 (125)\\
&  4 & 0.52$\pm$0.01 & 129$\pm$2 & 0.47$\pm$0.01 & 0.81 (126)	& 0.55$\pm$0.01 & 130$\pm$2 & 0.79 (125)\\
\hline
1997.5.1 &  1 & 0.45$\pm$0.03 & 129$\pm$4 & 0.45$\pm$0.02 & 1.11 (126)	& 0.48$\pm$0.03 & 131$\pm$6 & 1.11 (125)\\
&  2 & 0.57$\pm$0.03 & 144$\pm$3 & 0.44$\pm$0.02 & 1.03 (126)	& 0.59$\pm$0.03 & 146$\pm$4 & 1.03 (125)\\
& 3 & 0.58$\pm$0.02 & 178$\pm$4 & 0.49$\pm$0.02 & 0.84 (126)	& 0.61$\pm$0.03 & 170$\pm$5 & 0.85 (125)\\
& 4 & 0.52$\pm$0.01 & 143$\pm$2 & 0.44$\pm$0.01 & 1.03 (126)	& 0.55$\pm$0.02 & 145$\pm$2 & 1.01 (125)\\
\hline
1997.5.2& 1 & 0.57$\pm$0.02 & 171$\pm$3 & 0.46$\pm$0.01 & 0.86 (126)	& 0.60$\pm$0.02 & 174$\pm$4 & 0.85 (125)\\
\hline
1997.5.3 & 1 & 0.42$\pm$0.03 & 120$\pm$4 & 0.48$\pm$0.02 & 0.97 (126)	& 0.39$\pm$0.06 & 116$\pm$4 & 0.96 (125)\\
& 2 & 0.42$\pm$0.02 & 123$\pm$3 & 0.44$\pm$0.02 & 1.13 (126)	& 0.41$\pm$0.05 & 120$\pm$4 & 1.14 (125)\\
\hline
1997.5.4 & 1 & 0.31$\pm$0.02 & 105$\pm$4 & 0.50$\pm$0.03 & 1.23 (126)	& 0.22$\pm$0.09 & 104$\pm$5 & 1.25 (125)\\
& 2 & 0.32$\pm$0.02 & 103$\pm$3 & 0.50$\pm$0.02 & 1.23 (126)	& 0.32$\pm$0.05 & 103$\pm$4 & 1.24 (125)\\
\hline
1997.5.5& 1 & 0.38$\pm$0.02 & 127$\pm$4 & 0.47$\pm$0.02 & 1.24 (126)	& 0.42$\pm$0.03 & 131$\pm$5 & 1.24 (125)\\
& 2 & 0.37$\pm$0.02 & 127$\pm$3 & 0.46$\pm$0.02 & 1.24 (126)	& 0.40$\pm$0.03 & 129$\pm$4 & 1.25 (125)\\
\hline
1998.4.21- &  1 & 0.69$\pm$0.04 & 341$\pm$8 & 0.42$\pm$0.02 & 1.04 (126)	& 0.69$\pm$0.04 & 357$\pm$13 & 1.01 (125)\\
 &  2 & 0.85$\pm$0.02 & 404$\pm$6 & 0.42$\pm$0.01 & 1.12 (126)	& 0.93$\pm$0.03 & 397$\pm$7 & 1.06 (125)\\
&  3 & 0.87$\pm$0.02 & 378$\pm$5 & 0.37$\pm$0.01 & 1.03 (126)	& 0.91$\pm$0.03 & 379$\pm$6 & 1.04 (125)\\
&  4 & 0.83$\pm$0.03 & 337$\pm$5 & 0.37$\pm$0.01 & 0.93 (126)	& 0.87$\pm$0.04 & 337$\pm$7 & 0.94 (125)\\
&  5 & 0.74$\pm$0.03 & 300$\pm$5 & 0.34$\pm$0.01 & 0.90 (126)	& 0.72$\pm$0.04 & 310$\pm$7 & 0.86 (125)\\
&  6 & 0.78$\pm$0.03 & 274$\pm$4 & 0.37$\pm$0.01 & 1.31 (126)	& 0.82$\pm$0.03 & 274$\pm$5 & 1.33 (125)\\
&  7 & 0.64$\pm$0.03 & 239$\pm$4 & 0.33$\pm$0.02 & 1.04 (126)	& 0.68$\pm$0.03 & 235$\pm$5 & 1.02 (125)\\
&  8 & 0.67$\pm$0.03 & 236$\pm$5 & 0.38$\pm$0.02 & 0.95 (126)	& 0.70$\pm$0.04 & 235$\pm$6 & 0.96 (125)\\
\hline
1998.4.23- &  1 & 0.65$\pm$0.03 & 309$\pm$8 & 0.44$\pm$0.02 & 1.11 (126)	& 0.70$\pm$0.05 & 287$\pm$8 & 1.06 (125)\\
&  2 & 0.58$\pm$0.03 & 257$\pm$6 & 0.38$\pm$0.02 & 1.07 (126)	& 0.62$\pm$0.03 & 257$\pm$8 & 1.08 (125)\\
&  3 & 0.56$\pm$0.02 & 247$\pm$5 & 0.40$\pm$0.02 & 0.90 (126)	& 0.58$\pm$0.02 & 254$\pm$7 & 0.88 (125)\\
&  4 & 0.58$\pm$0.03 & 248$\pm$6 & 0.43$\pm$0.02 & 1.04 (126)	& 0.61$\pm$0.03 & 239$\pm$7 & 1.01 (125)\\
&  5 & 0.57$\pm$0.02 & 253$\pm$5 & 0.44$\pm$0.02 & 0.86 (126)	& 0.60$\pm$0.02 & 251$\pm$7 & 0.87 (125)\\
&  6 & 0.52$\pm$0.02 & 233$\pm$5 & 0.44$\pm$0.02 & 1.17 (126)	& 0.53$\pm$0.03 & 227$\pm$6 & 1.18 (125)\\
&  7 & 0.49$\pm$0.02 & 220$\pm$5 & 0.43$\pm$0.02 & 1.34 (126)	& 0.51$\pm$0.03 & 218$\pm$6 & 1.36 (125)\\
&  8 & 0.46$\pm$0.03 & 219$\pm$7 & 0.38$\pm$0.02 & 1.07 (126)	& 0.44$\pm$0.06 & 213$\pm$8 & 1.08 (125)\\
&  9 & 0.45$\pm$0.03 & 217$\pm$7 & 0.42$\pm$0.02 & 1.04 (126)	& 0.34$\pm$0.09 & 203$\pm$7 & 1.01 (125)\\
& 10 & 0.44$\pm$0.03 & 203$\pm$5 & 0.40$\pm$0.02 & 1.14 (126)	& 0.39$\pm$0.06 & 194$\pm$6 & 1.14 (125)\\
\enddata
\tablenotetext{a}{Each number represents the
observation divided into 10 ksec segments}
\tablenotetext{b}{in keV}
\tablenotetext{c}{in $10^{-12}$\flux}
\tablenotetext{d}{Definition is given in Table 1}
\tablecomments{All errors represent the 1$\sigma$ error.}
\end{deluxetable}

\begin{deluxetable}{lc | cc| c c cc}
\tablecaption{Simultaneous TeV flux and Synchrotron Peak Energy 
and Flux for Mrk~421
\label{tbl:xtev}}
\tablewidth{0pt}
\tablecolumns{9} 
\tabletypesize{\scriptsize}
\tablehead{
Start Time & Duration  &
\multicolumn{2}{c}{TeV}  &
\multicolumn{4}{c}{X--ray}  \\
(MJD)   & (ksec)  &
Observatory & TeV flux\tablenotemark{a} &
$\ep$\tablenotemark{b} & $\nfnp$\tablenotemark{c} &
$a$\tablenotemark{d} &$\chi^2_{\nu}$ (dof)
}
\startdata
50926.8814 & 11.0 &HEGRA&  0.46$\pm$0.10& 0.48$\pm$0.03 & 226$\pm$6 & 0.45$\pm$0.04 & 0.75(33)\\ 
50927.1390 & 13.7&\whipple&0.33$\pm$0.09& 0.47$\pm$0.01 & 218$\pm$2 & 0.42$\pm$0.01 & 0.95(216)\\
50927.8839 & 10.5 &HEGRA&  0.79$\pm$0.12& 0.74$\pm$0.02 & 381$\pm$3 & 0.43$\pm$0.01 & 1.18(210)\\ 
50928.2273 & 5.3 &\whipple&0.33$\pm$0.07& 0.56$\pm$0.02 & 269$\pm$3 & 0.42$\pm$0.01 & 0.95(142)\\
50928.8824 & 10.5 &HEGRA&  0.58$\pm$0.10& 0.66$\pm$0.02 & 316$\pm$2 & 0.41$\pm$0.01 & 0.90(221)\\ 
50929.8848 & 10.1 &HEGRA&  1.10$\pm$0.13& 0.77$\pm$0.02 & 424$\pm$2 & 0.36$\pm$0.01 & 1.13(205)\\ 
50930.1516 & 12.2&\whipple&0.54$\pm$0.05& 0.81$\pm$0.02 & 364$\pm$2 & 0.40$\pm$0.01 & 0.99(274)\\
50930.8845 & 9.8  &HEGRA&  1.41$\pm$0.15& 0.66$\pm$0.03 & 472$\pm$4 & 0.39$\pm$0.01 & 1.15(244)\\ 
50931.1467 & 10.3&\whipple&0.58$\pm$0.14& 0.72$\pm$0.02 & 367$\pm$3 & 0.38$\pm$0.01 & 0.94(229)\\
50932.2041 & 6.9 &\whipple&0.67$\pm$0.07& 0.84$\pm$0.02 & 367$\pm$2 & 0.34$\pm$0.01 & 1.31(239)\\
50933.2358 & 3.4 &\whipple&0.67$\pm$0.11& 0.90$\pm$0.02 & 454$\pm$2 & 0.38$\pm$0.01 & 1.21(238)\\
\enddata
\tablenotetext{a}{in Crab units}
\tablenotetext{b}{in keV}
\tablenotetext{c}{in $10^{-12}$\flux}
\tablenotetext{d}{Definition is given in Table 1}
\tablecomments{All errors represent the 1$\sigma$ error.}
\end{deluxetable}

\begin{figure}
\epsscale{0.4}
\plotone{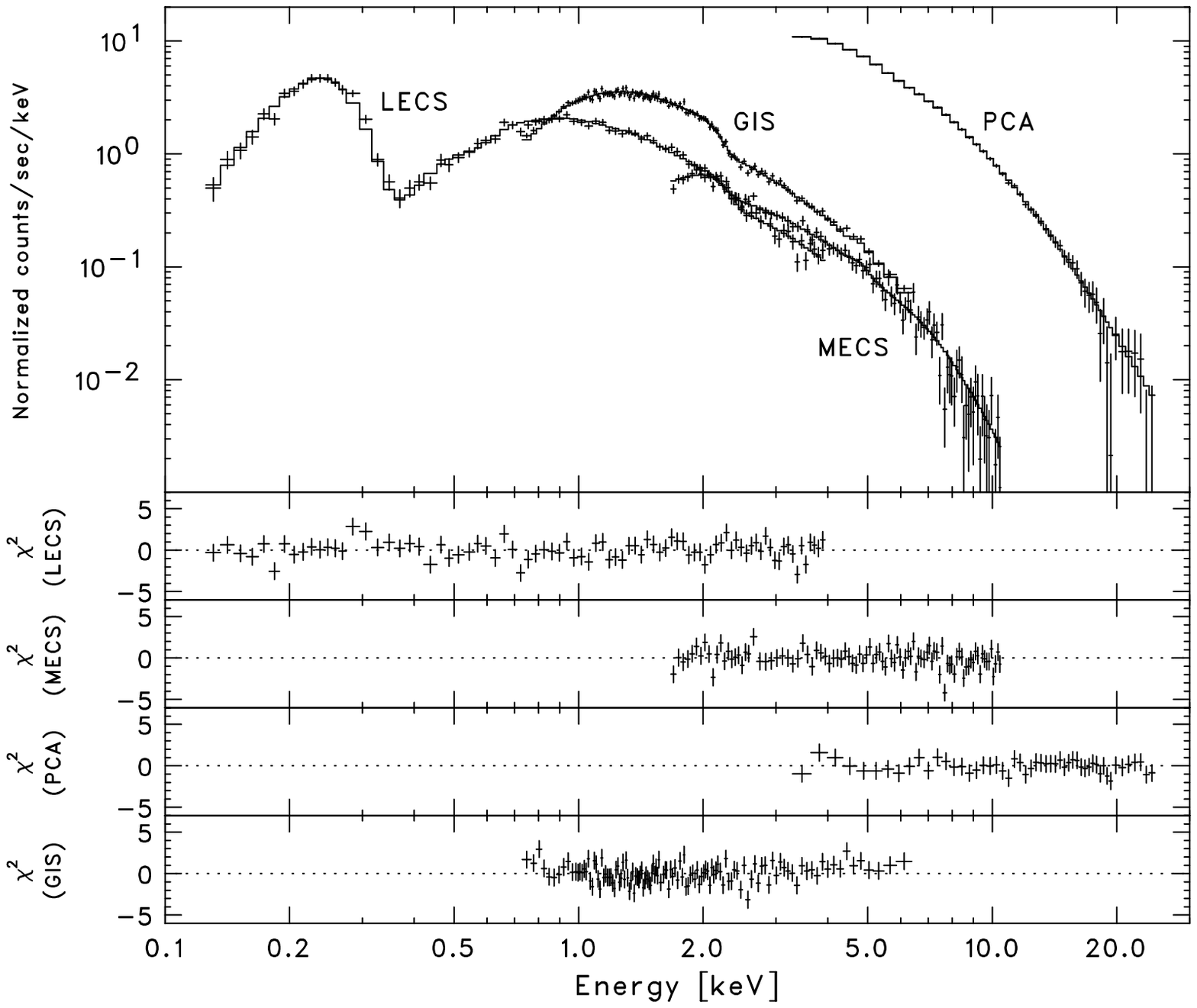}
\end{figure}
\begin{figure}
\epsscale{0.4}
\plotone{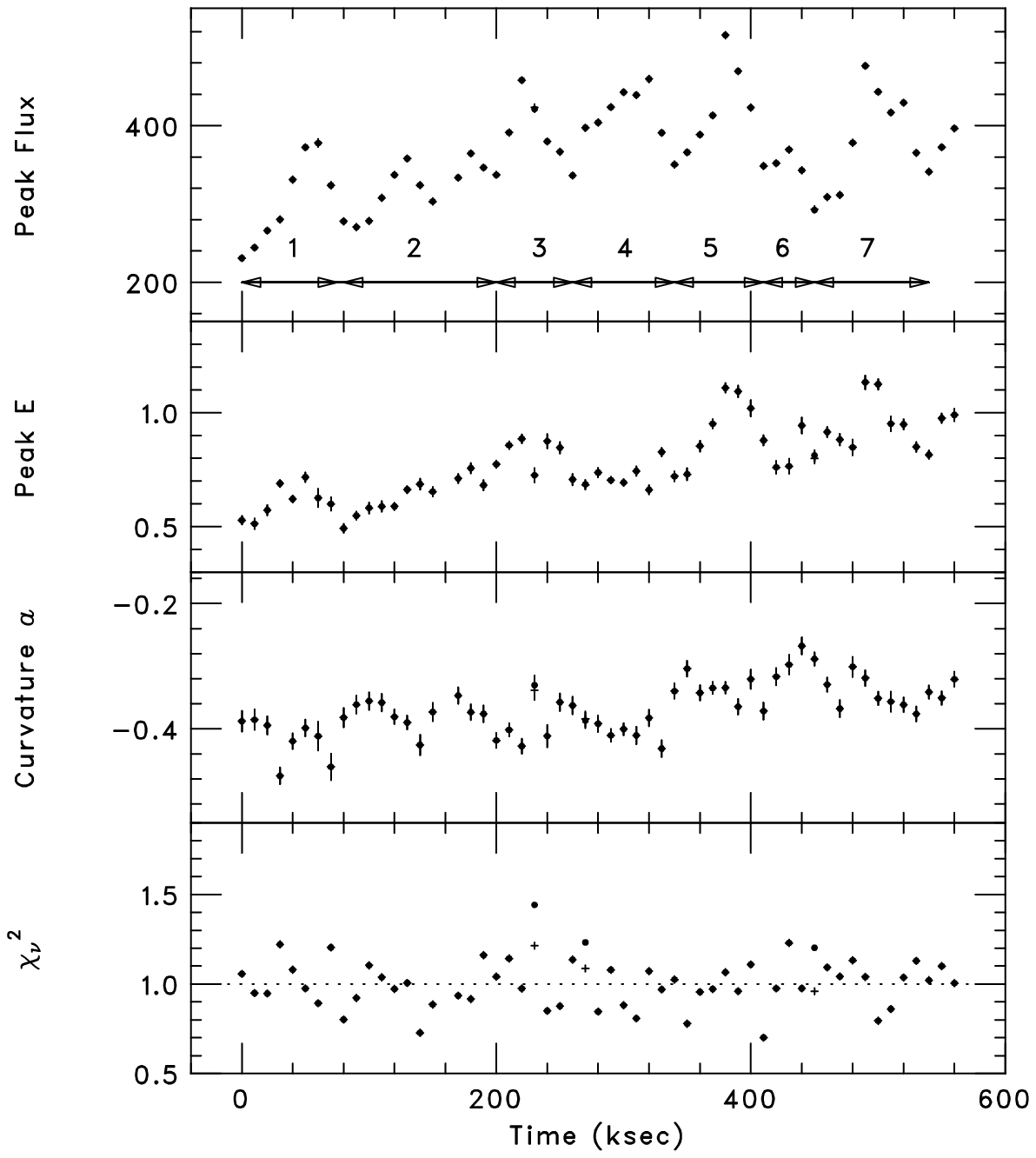}
\end{figure}
\begin{figure}
\epsscale{0.4}
\plotone{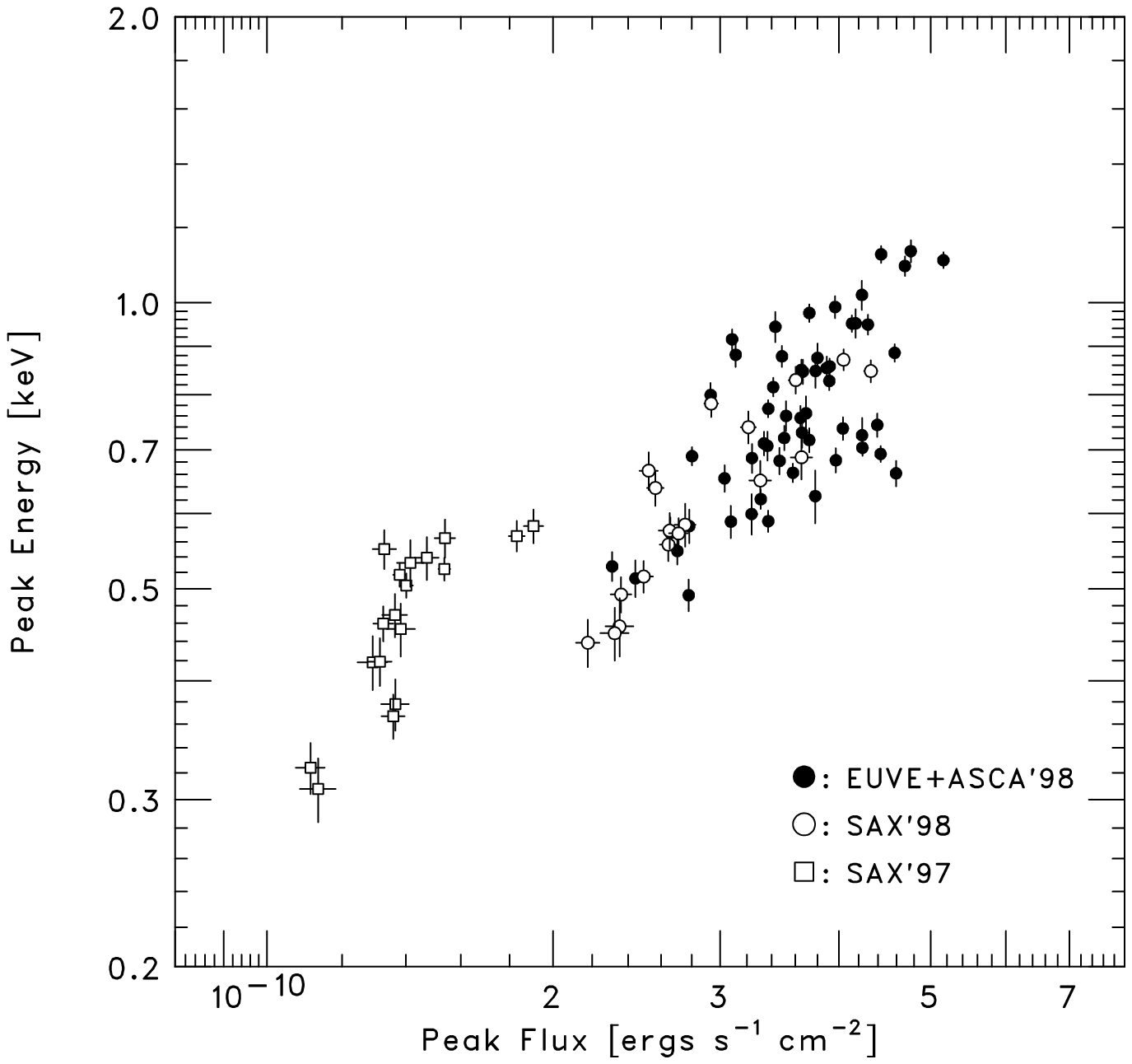}
\end{figure}
\begin{figure}
\epsscale{0.6}
\plotone{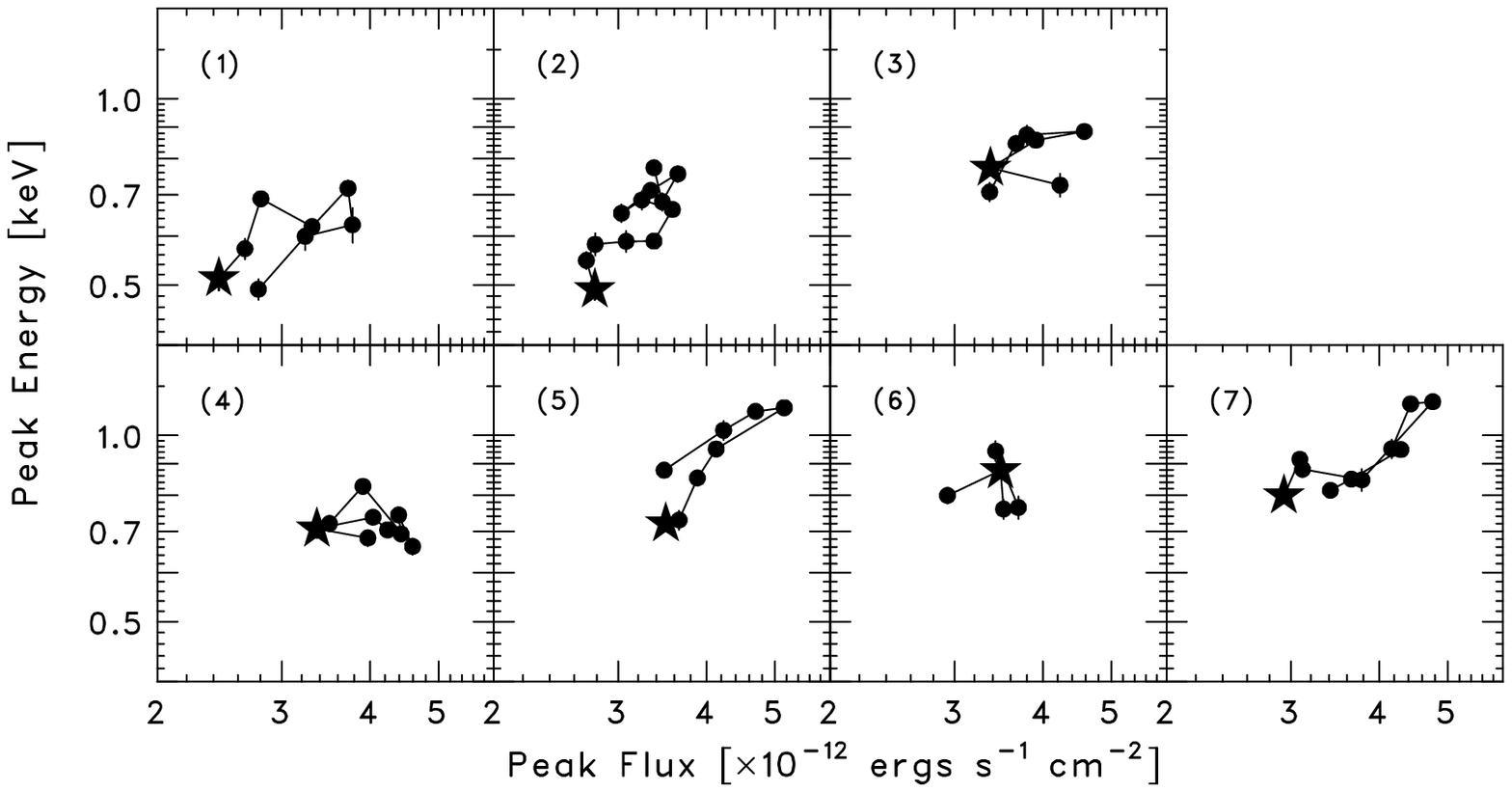}
\end{figure}
\begin{figure}
\epsscale{0.3}
\plotone{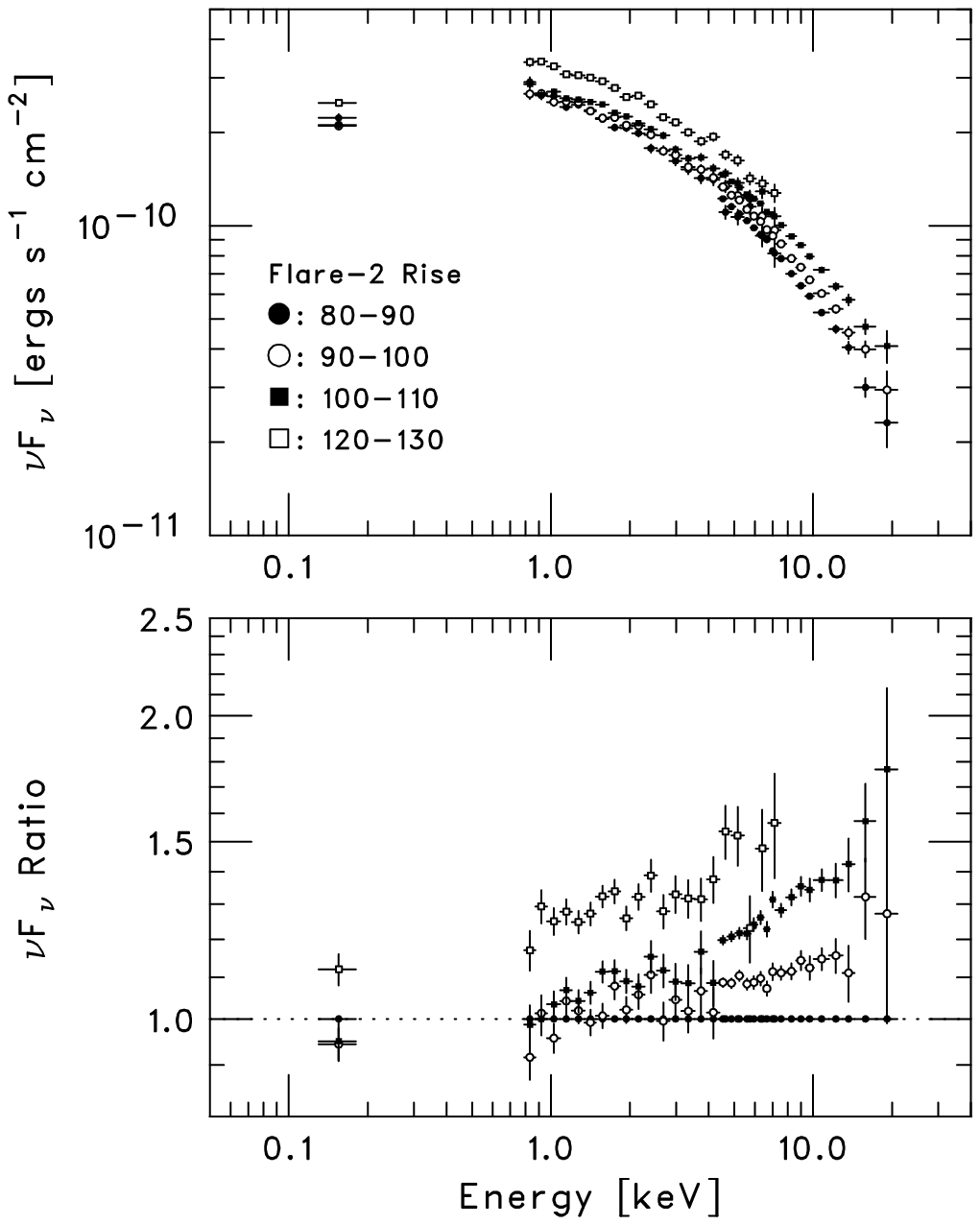}
\end{figure}
\begin{figure}
\epsscale{0.3}
\plotone{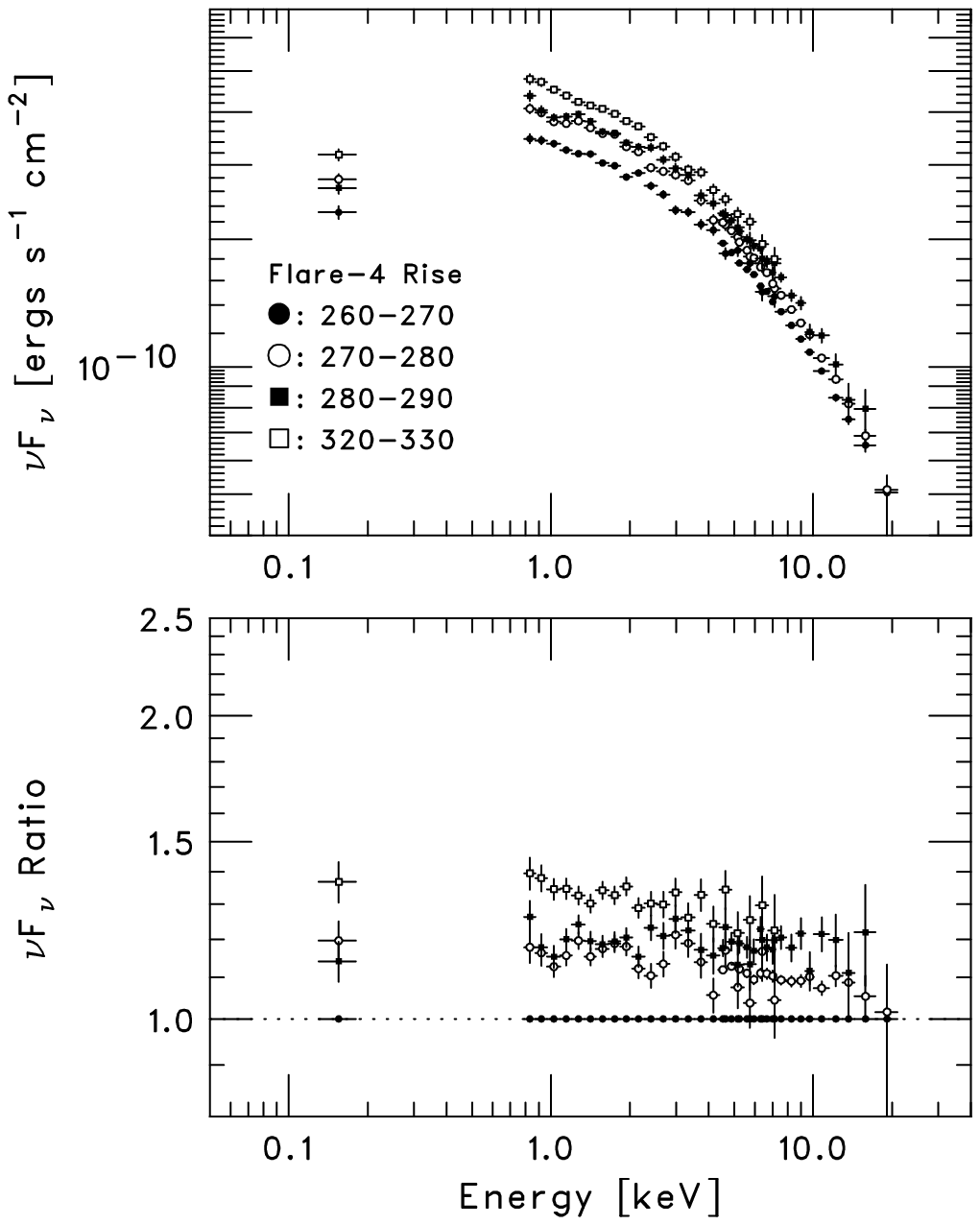}
\end{figure}
\begin{figure}
\epsscale{0.3}
\plotone{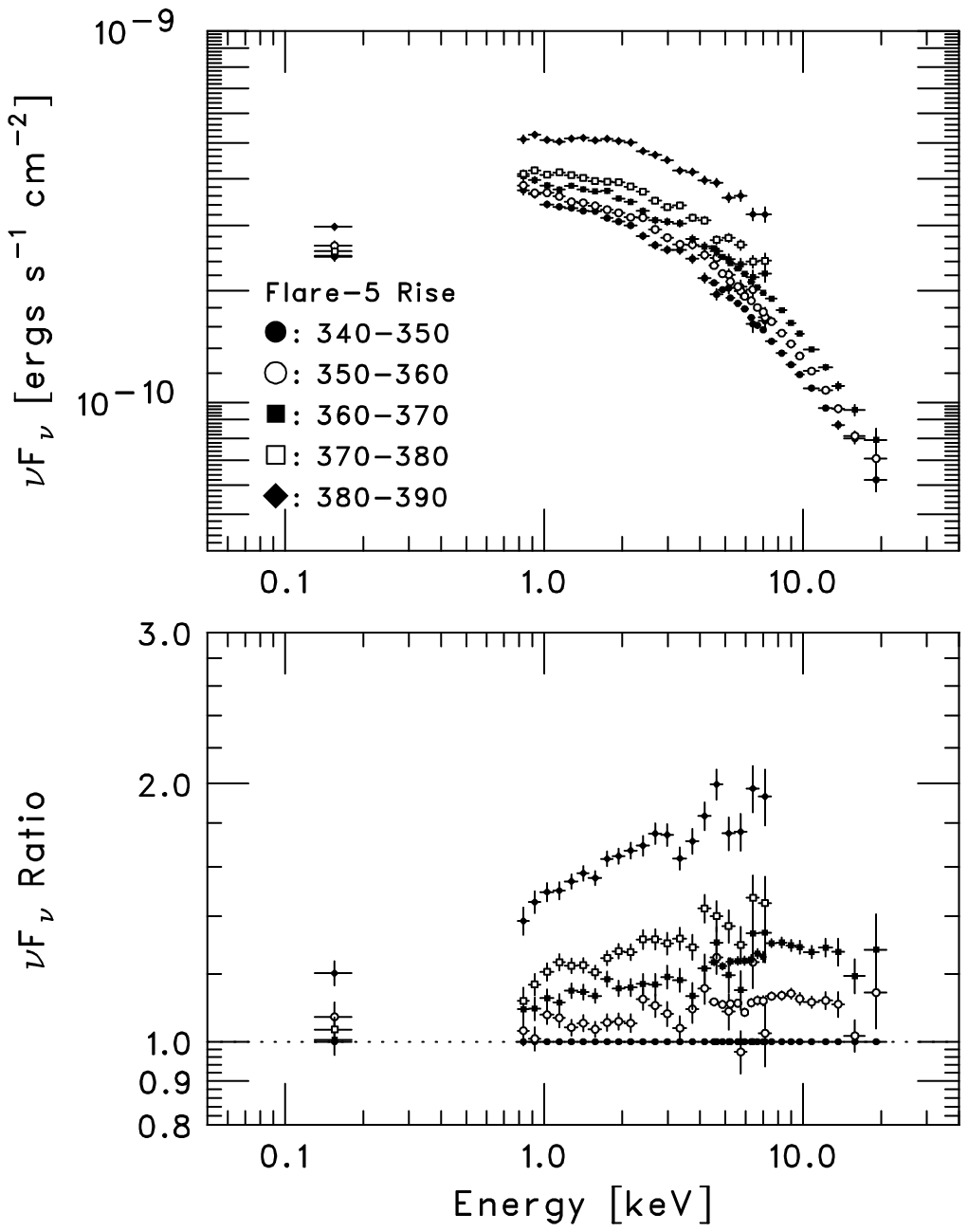}
\end{figure}
\begin{figure}
\epsscale{0.4}
\plotone{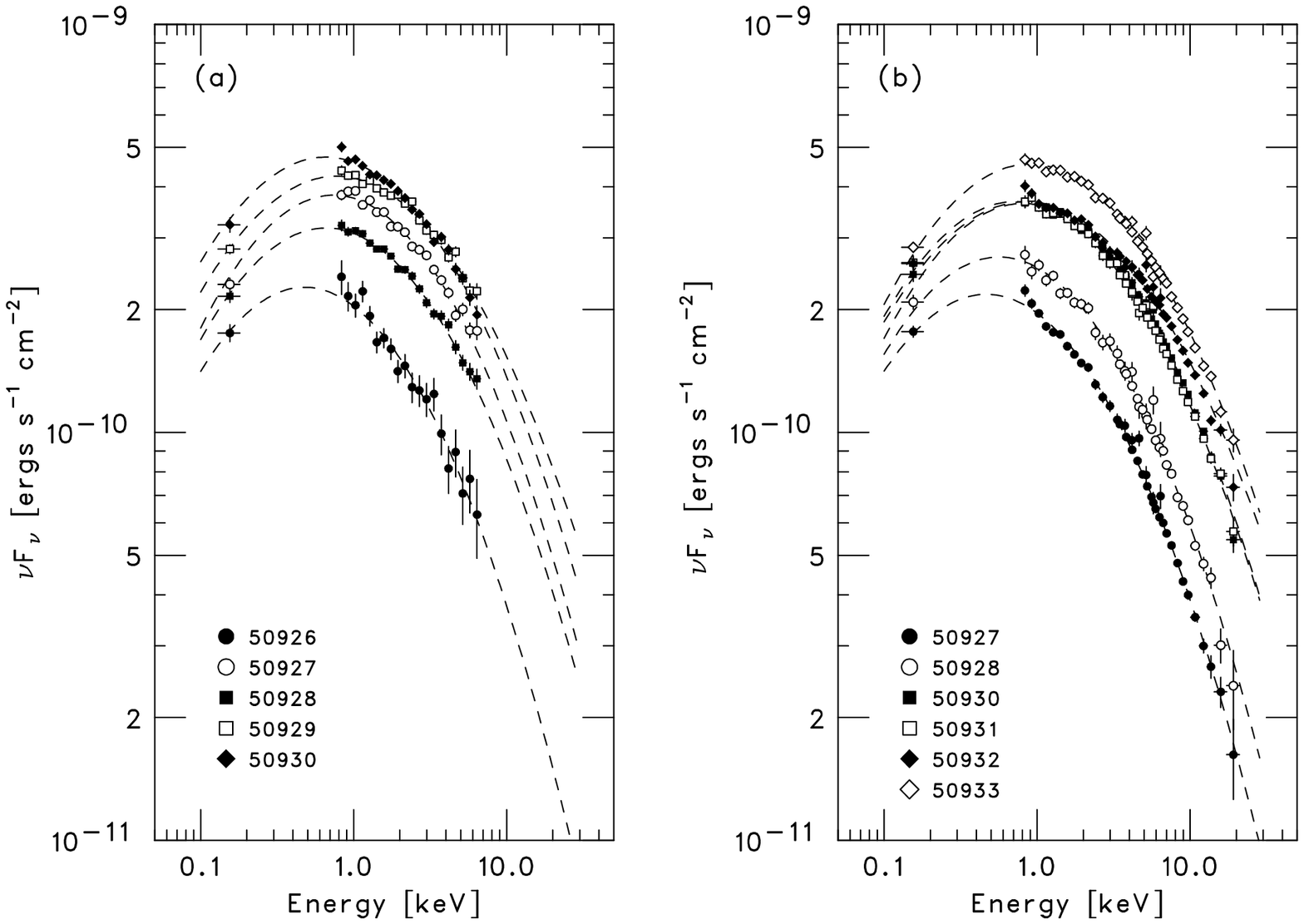}
\end{figure}
\begin{figure}
\epsscale{0.4}
\plotone{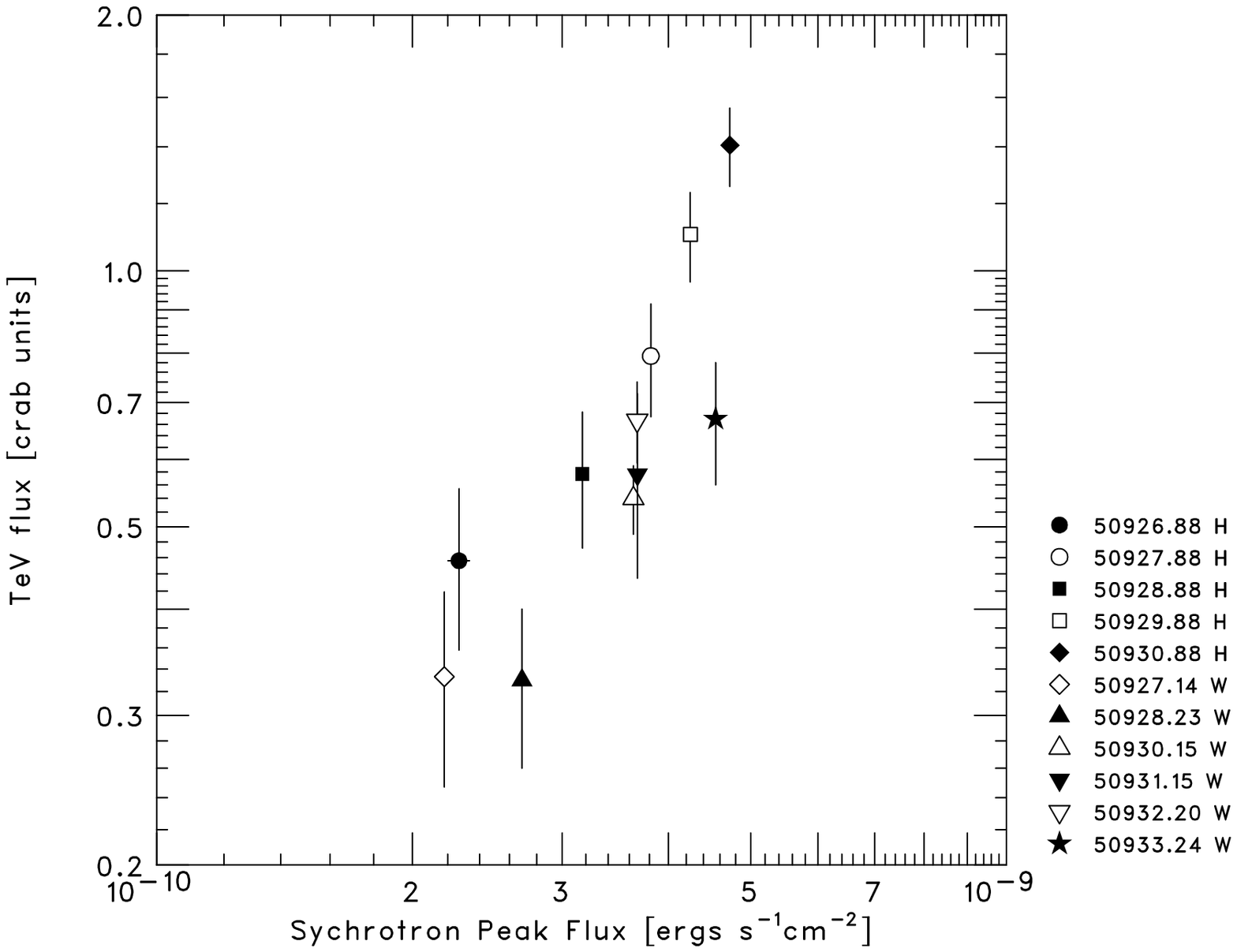}
\end{figure}

\end{document}